\documentstyle[aps]{revtex}
\begin{document}
\voffset 0.8truecm
\title{A note on quantum entropy inequalities and channel capacities 
}

\author{
Heng Fan
}
\address{
Quantum computation and quantum information project,
ERATO, \\
Japan Science and Technology Corporation,\\
Daini Hongo White Building 201,
Hongo 5-28-3, Bunkyo-ku, Tokyo 133-0033, Japan.
}
\maketitle

\begin{abstract}
Quantum entropy inequalities are studied.
Some quantum entropy inequalities are obtained by several methods.
For entanglement breaking channel, we show that
the entanglement-assisted classical capacity is
upper bounded by $\log d$.
A relationship between
entanglement-assisted and one-shot unassisted capacities
is obtained. This relationship shows the entanglement-assisted
channel capacity is upper bounded by the sum of $\log d$ and
the one-shot unassisted classical capacity.
\end{abstract}

\section{Introduction}
Quantum information theory has been attracting a great deal
of interests. Several capacities of quantum channels
are proposed and studied, such as the
Holevo-Schumacher-Westmoreland channel capacity\cite{H0,SW} and
the recent proposed entanglement-assisted classical
capacity\cite{BSST1,BSST2},
adaptive classical capacity\cite{S0}.
In studying capacities of quantum channels,
the quantum entropy inequalities are very important.
In ref.\cite{NC,We,Ruskai,CA,BNS,HW}, a survey of
quantum entropy inequalities are presented.                                      
Some of these quantum inequalities are independent but
equivalent, i.e. they are necessary and sufficient
conditions to each other\cite{Ruskai}.
In some cases, the results can be obtained
much easier from one quantum entropy inequality
than from others. So, all of these inequalities are necessary.
It will be better if we can find more inequalities.
In this paper, we try to study some of these quantum entropy
inequalities and to find their applications in channel
capacities. In particular, we propose the strong
concavity of von Neumann entropy inequality and prove it
by several methods.

The additivity of classical capacity of quantum channels is one of the
fundamental problems in studying the quantum information
theory. The additivity of classical capacity
of several special channels is proved, such as unital qubit
channels\cite{King1}, depolarizing
channels\cite{King2} and entanglement breaking
channels\cite{S}.
By using directly the strong concavity
of von Neumann entropy, we give a simple proof of
the additivity of classical
capacity of entanglement breaking channels.

It is known that the classical capacity of quantum channels may be
enhanced with prior entanglement such as the super-dense
coding protocol\cite{BW}. A general theorem called
entanglement-assisted classical capacity
was proposed and proved recently
concerning about the classical capacity of quantum channels
with the help of shared entanglement between the sender and
the receiver\cite{BSST1,BSST2}.
It can be expected that if the channel itself
is entanglement breaking, its entanglement-assisted classical
capacity has less advantage than other channels.
Really, we show in this paper for entanglement breaking channel,
the entanglement-assisted classical capacity is upper bounded
by $\log d$ while generally we have an extra term $\chi $.

A simple proof of the entanglement-assisted channel
capacity was given by Holevo\cite{H}, he also found
the entanglement-assisted channel capacity is upper
bounded by the sum of $\log d$ and the unassisted
classical capacity. We shall show further in this paper that
the entanglement-assisted channel capacity is upper
bounded by the sum of $\log d$ and the one-shot
unassisted classical capacity. This result also eliminates
one of the possible ways in which one might prove
non-additivity of the classical capacity.

\section{Equivalent quantum entropy inequalities}
There are 4 equivalent quantum entropy inequalities as
reviewed by Ruskai\cite{Ruskai}. In this section, we point out 
that we can add another equivalent entropy
inequality.

First, let us introduce some definitions.
The von Neumann entropy is defined as:
\begin{eqnarray}
S\left( \rho \right) \equiv -{\rm Tr}(\rho\log \rho ), 
\end{eqnarray}
where $\rho $ is the density operator. The relative entropy
is defined as:
\begin{eqnarray}
S\left( \rho ||\sigma \right)
\equiv {\rm Tr}\rho ({\rm log}\rho -{\rm log}\sigma ).
\end{eqnarray}

In a recent review paper, Ruskai listed
the first 4
equivalent quantum entropy inequalities as presented
as follows, see \cite{Ruskai} and the references therein:

\noindent
1, Monotonicity of relative entropy under completely positive, trace
preserving maps:
\begin{eqnarray}
S\left( \Phi (\rho )||\Phi (\sigma )\right)\le S\left( \rho ||\sigma \right).
\end{eqnarray}
2, Monotonicity of relative entropy under partial trace:
\begin{eqnarray}
S\left( \rho _A||\sigma _A\right)\le S\left( \rho _{AC}||
\sigma _{AC} \right).
\end{eqnarray}
3, Strong subadditivity of von Neumann entropy I and II, where
I and II are equivalent:
\begin{eqnarray}
&&I),~~ S\left( \rho _A\right) +S\left( \rho _B\right)
\le S\left( \rho _{AC}\right) +S\left( \rho _{BC}\right);
\nonumber \\
&&II),~~ S\left( \rho _{ABC}\right) +S\left( \rho _B\right)
\le S\left( \rho _{AB}\right) +S\left( \rho _{BC}\right).
\end{eqnarray}
4, Joint convexity of relative entropy:
\begin{eqnarray}
S\left( \sum _ip_i\rho ^i||\sum _ip_i\sigma ^i\right)
\le \sum _ip_iS\left( \rho ^i||\sigma ^i\right) .
\end{eqnarray}
5, Actually, we can add another equivalent inequality,
concavity of conditional entropy:
\begin{eqnarray}
S\left( \sum _ip_i\rho _{AB}^i\right)
-S\left( \sum _ip_i\rho _{B}^i\right)
\ge \sum _ip_i[S\left( \rho _{AB}^i\right)
-S\left( \rho _B^i\right) ].
\end{eqnarray}
The last inequality was deduced from 4, joint convexity of relative
entropy in Ref.\cite{NC}. Then it was used to deduce the inequality
3, strong subadditivity. So, inequality 5, the concavity of
conditional entropy is an equivalent inequality with the other 4
inequalites.

In the textbook of Nielsen and Chuang\cite{NC}, inequality 5 is obtained
from 4, the joint convexity. 
Next, we show two other methods to obtain the concavity of conditional
entropy. First, we use 2, monotonicity of relative entropy under partial
trace.
Suppose $\rho _{AB}=\sum _ip_i\rho _{AB}^i$, so
$\rho _B=\sum _ip_i\rho _B^i$.
Since inequality 2, we have
\begin{eqnarray}
S\left( \rho _B^i||\rho _B\right) \le S\left( \rho _{AB}^i||\rho _{AB}
\right) .
\end{eqnarray}
So, the average of relative entropies has inequality
\begin{eqnarray}
\sum _ip_i
S\left( \rho _B^i||\rho _B\right) \le
\sum _ip_iS\left( \rho _{AB}^i||\rho _{AB}
\right) .
\end{eqnarray}
From the definition of relative entropy, we obtain 5,
\begin{eqnarray}
S\left( \rho _{AB}\right) -S\left( \rho _B\right)
\ge \sum _ip_i[S\left( \rho _{AB}^i\right) -
S\left( \rho _B^i\right) ].
\end{eqnarray}
Secondly, we also use the joint
convexity to deduce 5, but by a different method.
The joint convexity of relative entropy means,
\begin{eqnarray}
S\left( \sum _ip_i\rho _{AB}^i||\sum _ip_i\rho _B^i\right)
\le \sum _ip_iS\left( \rho _{AB}^i||\rho _B^i\right) .
\end{eqnarray}
By definition, we have
\begin{eqnarray}
-S\left( \rho _{AB}\right) +S\left( \rho _B\right)
\le -\sum _ip_i[S\left( \rho _{AB}^i\right)
-S\left( \rho _B^i\right) .
\end{eqnarray}
This is exactly 5, the concavity of conditional entropy.

Since these 5 inequalities are equivalent, we can obtain any of them
by one of the other 4 inequalities.
Recently, Bennett et al proposed and proved the
entanglement-assisted channel capacity\cite{BSST1,BSST2}.
Holevo subsequently
gave a modified proof\cite{H}, and one of the simplifications
is due to the replacement of strong subadditivity
by concavity of conditional entropy, i.e., the fifth
inequality was used directly in Ref.\cite{H} rather than the third
inequality used in Ref.\cite{BSST2} though
they are equivalent.

\section{Strong concavity of von Neumann entropy}
In this section, we propose the following
quantum entropy inequality: strong concavity of von
Neumann entropy,
\begin{eqnarray}
S\left( \sum _ip_i\rho _A^i\otimes \rho _B^i\right)
\ge \max \{ \sum _ip_iS\left( \rho _A^i\right)
+S\left( \sum _ip_i\rho _{B}^i\right),
\sum _ip_iS\left( \rho _B^i\right)
+S\left( \sum _ip_i\rho _{A}^i\right) \} .
\label{sc}
\end{eqnarray}
To prove this inequality, we need to show that
both of the following two inequalities hold,
\begin{eqnarray}
S\left( \sum _ip_i\rho _A^i\otimes \rho _B^i\right)
\ge \sum _ip_iS\left( \rho _A^i\right)
+S\left( \sum _ip_i\rho _{B}^i\right) .
\end{eqnarray}
and 
\begin{eqnarray}
S\left( \sum _ip_i\rho _A^i\otimes \rho _B^i\right)
\ge \sum _ip_iS\left( \rho _B^i\right)
+S\left( \sum _ip_i\rho _{A}^i\right) .
\end{eqnarray}
We denote $\rho _A=\sum _ip_i\rho _A^i, \rho _B=\sum _ip_i\rho _B^i$.

In the following, we present several methods to derive the
strong concavity of quantum entropy.

\noindent 
A, Due to 2, monotonicity of relative entropy, we have
\begin{eqnarray}
S\left( \rho _A^i||\rho _A\right)
\le S\left( \rho _A^i\otimes \rho _B^i||\sum _ip_i\rho _A^i\otimes
\rho _B^i\right) .
\end{eqnarray}
Take average with probability $p_i$, we have
\begin{eqnarray}
\sum _ip_iS\left( \rho _A^i||\rho _A\right)
\le \sum _ip_iS\left( \rho _A^i\otimes \rho _B^i||\sum _jp_j\rho _A^j\otimes
\rho _B^j\right) .
\end{eqnarray}
So, we have
\begin{eqnarray}
-\sum _ip_iS\left( \rho _A^i\right)
+S\left( \rho _A\right)
\le -\sum _ip_i[S\left( \rho _A^i\right)
+S\left( \rho _B^i\right) ]
+S\left( \sum _ip_i\rho _A^i\otimes \rho _B^i\right) .
\end{eqnarray}
Thus we obtain the strong concavity.

\noindent C, Due to 4, joint convexity of relative entropy,
\begin{eqnarray}
S\left( \sum _ip_i\rho _A^i\otimes \rho _B^i||
\sum _ip_i\rho _A^i\right)
&\le &\sum _ip_iS\left(
\rho _A^i\otimes \rho _B^i||\rho _A^i\right)
\nonumber \\
&=&
-\sum _ip_iS\left( \rho _B^i\right) .
\end{eqnarray}
We have the strong concavity of von Neumann entropy.

\noindent B, Due to 5, concavity of conditional entropy, we have
\begin{eqnarray}
S\left( \sum _ip_i\rho _A^i\otimes \rho _B^i\right)
-S\left( \sum _ip_i\rho _A^i\right)
&\ge &\sum _ip_i[S\left( \rho _{A}^i\otimes \rho _B^i\right)
-S\left( \rho _A^i\right )]
\nonumber \\
&=&\sum _ip_iS\left( \rho _B^i\right) .
\end{eqnarray}
Then we arrive at the strong concavity.
Since the situations for $\rho _A$ and $\rho _B$ are the same,
we know that Eq.(\ref{sc}) holds.
To the author's knowlege, this inequality has not been
explicit proposed previously.
Though it is implied in the study of the
additivity of entanglement breaking channel\cite{S}
and the additivity of entanglement of formation
in some cases\cite{F}.
Next, we would like to show some applications of this
inequality.

\section{Application of strong concavity in channel
capacity of entanglement breaking channel}
Recently,
Shor proved the additivity of the classical
capacity of the entanglement breaking quantum channel\cite{S}.
Both c-q (classical-quantum) and q-c (quantum-classical) channels
are special cases of entanglement breaking channels.
And the entanglement breaking channel can be expressed as a q-c-q
channel.
Other properties and conjectures of entanglement
breaking channel can be found in Ref.\cite{Ruskai2}.
We next give a simple proof of
the additivity of channel capacity of the entanglement
breaking channel by directly use the strong
concavity inequality though there are no essential differences from
Shor's original proof.

An entanglement breaking channel $\Phi $ means
$(I\otimes \Phi )\rho _{AB}$ is always a separable state, which
can be written as\cite{Werner},
\begin{eqnarray}
(I\otimes \Phi )\rho _{AB}
=\sum _ip_i\rho _A^i\otimes \rho _B^i.
\end{eqnarray}
So we know $\Phi (\rho _B)=\sum _ip_i\rho _B^i$.
Suppose $\sum _jq_j\rho _{AB}^j=\rho _{AB}$ are the
optimal signal states for channel $\Psi \otimes \Phi $,
where $\Psi $ is an arbitrary quantum channel. The
Holevo-Schumacher-Westmoreland channel capacity
$\chi ^*(\Psi \otimes \Phi ) $ takes the following form
\begin{eqnarray}
\chi ^*(\Psi \otimes \Phi )
&=&
\sum _jq_jS\left( (\Psi \otimes \Phi )(\rho ^j_{AB})
||(\Psi \otimes \Phi )(\sum _jq_j\rho _{AB}^j)\right)
\nonumber \\
&=&-\sum _jq_j
S\left( \sum _ip_{ji}\Psi (\rho _{A}^{ji})\otimes \rho _{B}^{ji}
\right)
+S\left( (\Psi \otimes \Phi )(\rho _{AB})\right) .
\end{eqnarray}
Then using the strong concavity inequality to the first term and
subadditivity to the second term, we have
\begin{eqnarray}
\chi ^*(\Psi \otimes \Phi )
&\le &-\sum _{ji}q_jp_{ji}
S\left( \Psi (\rho _{A}^{ji})\right)
-\sum _jq_jS\left( \sum _ip_{ji}\rho _{B}^{ji}\right)
+S\left( \Psi (\rho _{A})\right)+
S\left( \Phi (\rho _{B})\right) 
\nonumber \\
&=&
\sum _{ji}q_jp_{ji}S\left( \Psi (\rho _{A}^{ji})||
\Psi (\rho _{A})\right)
+\sum _jq_jS\left( \Phi (\rho _{B}^j)||
\Phi (\rho _{B})\right) 
\nonumber \\
&\le & 
\chi ^*(\Psi )+\chi ^*(\Phi ).
\end{eqnarray}
Since the classical capacity of quantum channel
is strong additive, we thus know the capacity of entanglement
breaking channel is additive,
\begin{eqnarray}
\chi ^*(\Psi \otimes \Phi )
=\chi ^*(\Psi )+\chi ^*(\Phi ).
\end{eqnarray}

\section{Application of strong concavity of von Neumann entropy
in entanglement-assisted channel capacity for an
entanglement breaking channel}
Recently, Bennett et al\cite{BSST1,BSST2} (BSST theorem)
proposed and proved the
entanglement-assisted channel capacity in terms of quantum
mutual information. 
Holevo\cite{H} then gave a simple proof.
The BSST theorem states that the classical capacity of
the entanglement-assisted channel is written as the form
\begin{eqnarray}
C_{E}(\Phi )={\max} _{\rho _A\in {\cal {H}}_{in}}
S(\rho _A)+S\left( \Phi (\rho _A)\right)
-S\left( (\Phi \otimes I)(|\Psi _{AB}\rangle \langle
\Psi _{AB}|)\right) ,
\label{BSST}
\end{eqnarray}
where $|\Psi _{AB}\rangle $ is a purification of $\rho _{A}$.

Holevo\cite{H} pointed out that
there is a relationship between the entanglement-assisted and
unassisted capacities,
\begin{eqnarray}
C_E(\Phi )\le C(\Phi )+{\rm log}d,
\label{hresult}
\end{eqnarray}
where $d$ is the dimension of the Hilbert space ${\cal {H}}_{in}$.
If the additivity of the classical capacity holds, we can
replace $C(\Phi )$ by one-shot classical capacity $\chi ^*(\Phi )$.
Since Shor\cite{S} already proved that the classical capacity
of the entanglement breaking channel is additive.
So, for entanglement breaking channel $\Phi $, we have
\begin{eqnarray}
C_E(\Phi )\le \chi ^*(\Phi )+{\rm log}d.
\label{relation}
\end{eqnarray}
Next, we show that a tighter upper bound can be obtained for
entanglement breaking channel.
Because $\Phi $ is an entanglement breaking channel,
so we have
\begin{eqnarray}
(\Phi \otimes I)(|\Psi _{AB}\rangle
\langle \Psi _{AB}|)=\sum _ip_i\rho _A^i\otimes \rho _{B}^i,
\end{eqnarray}
where both $\rho _A^i$ and $\rho _B^i$ are pure states.
By strong concavity of von Neumann entropy, we know
\begin{eqnarray}
S\left(
(\Phi \otimes I)(|\Psi _{AB}\rangle
\langle \Psi _{AB}|)\right)
\ge
\left\{ \begin{array}{l}
S(\sum _ip_i\rho _A^i)+\sum _ip_iS(\rho _{B}^i)=
S\left(\Phi (\rho _A)\right),\\
\sum _ip_iS(\rho _A^i)+S(\sum _ip_i\rho _{B}^i)=
S(\rho _B)=S(\rho _A),\end{array}
\right.
\end{eqnarray}
Substitute these relations to BSST theorem (\ref{BSST}), we have
\begin{eqnarray}
C_E(\Phi )&\le &
{\max} _{\rho _A\in {\cal {H}}_{in}}
S(\rho _A)\le {\rm log}d,\\
{\rm or,}
~~~C_E(\Phi )&\le &
{\max} _{\rho _A\in {\cal {H}}_{in}}
S\left(\Phi (\rho _A)\right) \le {\rm log}d.
\end{eqnarray}
So, we know for entanglement breaking channel, the
entanglement-assisted classical capacity has an upper bound
\begin{eqnarray}
C_E(\Phi )\le {\rm log}d.
\end{eqnarray}
Comparing this relation with the general relation (\ref{relation}),
we find that
the term $\chi ^*(\Phi )$ does not appear here though it is
not always zero.
So, we show there is an upper bound for $C_E(\Phi )$ when
$\Phi $ is an entanglement breaking channel.
It might be interpreted as,
since the channel itself is entanglement-breaking,
the prior entanglement may not help much to increase the
classical capacity.  

\section{Relationship between entanglement-assisted and
one-shot unassisted capacities}
As already pointed out in last section, Holevo\cite{H} found
entanglement-assisted channel capacity is upper bounded
by the sum of $\log d$ and the unassisted classical 
capacity as relation (\ref{hresult}).
If the classical channel capacity is additive which
is a long-standing conjecture, then we have the
inequality
\begin{eqnarray}
C_E(\Phi )\le \chi ^*(\Phi )+{\rm log}d.
\label{conjec}
\end{eqnarray}
For an arbitrary quantum channel $\Phi $, if this relation
does not hold, that means $C(\Phi )>\chi ^*(\Phi )$,
thus the additivity conjecture of classical channel
capacity does not hold. So, (\ref{conjec}) may
provide a criterion to test the additivity problem
of classical capacity. However, we show in this section,
relation (\ref{conjec}) always holds for an arbitrary quantum
channel $\Phi $.

We assume that $\rho _A$ have the following pure state decomposition
\begin{eqnarray}
\rho _A=\sum _jq_j|\tilde {\Psi }^j_A\rangle \langle \tilde {\Psi }^j_A|.
\end{eqnarray}
Using the same technique as that of Ref.\cite{S}, we define
\begin{eqnarray}
|\tilde {\Psi }_{ABC}\rangle =\sum _j
\sqrt{q_j}|\tilde {\Psi }^j_A\rangle |j\rangle _B|j\rangle _C.
\end{eqnarray}
So, we have
\begin{eqnarray}
(\Phi \otimes I_{BC})(|\tilde {\Psi }_{ABC}\rangle
\langle \tilde {\Psi }_{ABC}|)
=\sum _{jj'}\sqrt{q_jq_{j'}}\Phi (|\tilde {\Psi }^i_A\rangle
\langle \tilde {\Psi }^{j'}_A|)
\otimes |j\rangle _B\langle j'|\otimes
|j\rangle _C\langle j'|.
\end{eqnarray}
With the help of the quantum entropy inequality, we obtain
\begin{eqnarray}
S\left(
(\Phi \otimes I_{BC})(|\tilde {\Psi }_{ABC}\rangle
\langle \tilde {\Psi }_{ABC}|)\right)
&\ge &S\left( (\Phi \otimes I_B)(\tilde {\rho }_{AB})\right)
-S(\tilde {\rho }_C)
\nonumber \\
&=&
S\left(
\sum _jq_j\Phi (|\tilde {\Psi }^j_A\rangle
\langle \tilde {\Psi }^j_A|)\otimes |j\rangle _B\langle j|\right)
-S\left( \sum _jq_j|j\rangle _C\langle j|\right)
\nonumber \\
&=&\sum _jq_j
S\left(\Phi (|\tilde {\Psi }^j_A\rangle
\langle \tilde {\Psi }^j_A|)\right) .
\end{eqnarray}
We know
\begin{eqnarray}
S\left( (\Phi \otimes I)(|\Psi _{AB}\rangle
\langle \Psi _{AB}|)\right)
=S\left( (\Phi \otimes I)(|\tilde {\Psi }_{ABC}\rangle
\langle \tilde {\Psi }_{ABC}|)\right) ,
\end{eqnarray}
where both $|\Psi _{AB}\rangle $ and
$|\tilde {\Psi }_{ABC}\rangle $ are purifications of $\rho _A$.
From BSST theorem (\ref{BSST}), we have
\begin{eqnarray}
C_{E}(\Phi )&=&{\max} _{\rho _A\in {\cal {H}}_{in}}
S(\rho _A)+S\left( \Phi (\rho _A)\right)
-S\left( (\Phi \otimes I)(|\Psi _{AB}\rangle \langle
\Psi _{AB}|)\right)
\nonumber \\
&\le &
{\max} _{\rho _A\in {\cal {H}}_{in}}
S(\rho _A)+S\left( \Phi (\rho _A)\right)
-\sum _jq_jS\left(\Phi(|\tilde {\Psi }^j_A\rangle \langle
\tilde {\Psi }^j_A|)\right)
\nonumber \\
&\le &
\log d+\chi ^*(\Phi ).
\end{eqnarray}
Thus, we conclude, for an arbitrary quantum channel
$\Phi $,  the entanglement-assisted
and one-shot unassisted capacities have the relationship
\begin{eqnarray}
C_E(\Phi )\le \chi ^*(\Phi )+{\rm log}d.
\end{eqnarray}
If the additivity of classical capacity holds, this relation
is the same as the relation (\ref{hresult}).
If the additivity does not hold for classical capacity,
this relation is tighter than (\ref{hresult}).

\section{Summary}
In summary, we pointed out that there are another
quantum entropy inequality, the concavity of conditional
entropy inequality,  to be equivalent to other
4 equivalent quantum entropy inequalities.
We proposed the strong concavity of von Neumann entropy and
proved it by several methods. Using directly this inequality,
the additivity of capacity of entanglement breaking channels
can be proved simply. We also showed for entanglement breaking
channel, the entanglement-assisted channel capacity is
upper bounded by $\log d$ which is tighter than the
general case. A new upper bound is obtained for the 
entanglement-assisted classical capacity,
the entanglement-assisted classical capacity is upper bounded
by the sum of $\log d$ and the one-shot unassisted capacity.
This result also eliminates one possible way to test the
non-additivity of classical capacity.

{\it Acknowlegements}: The author would like to thank
useful discussions with members of ERATO project.

\end{document}